\documentclass[10pt,a4paper,final,journal]{IEEEtran}

\usepackage[T1]{fontenc}
\usepackage[latin9]{inputenc}
\usepackage[active]{srcltx}
\usepackage{array}
\usepackage{multirow}
\usepackage{blindtext}
\usepackage{multicol}
\usepackage{amsmath}
\usepackage{amssymb}
\providecommand{\keywords}[1]{\textbf{\textit{Index terms}} }
\usepackage{graphicx}
\usepackage{setspace}
\usepackage{lipsum}
\usepackage[T1]{fontenc}
\usepackage{authblk}
\usepackage{url}
\usepackage{subfigure}
\usepackage{algorithm}
\usepackage{float}
\restylefloat{table}
\usepackage{xcolor}
\usepackage[noend]{algpseudocode}
\renewenvironment{IEEEbiography}[1]
  {\IEEEbiographynophoto{#1}}
  {\endIEEEbiographynophoto}
\makeatletter

\usepackage{cite}\usepackage{threeparttable}

\usepackage{picinpar}
\usepackage{rotating}

\usepackage{psfrag}\usepackage{placeins}\usepackage{times}
\usepackage{amsfonts}\usepackage{amsfonts}
\usepackage{epstopdf}

\newcommand*{\rom}[1]{\expandafter\@slowromancap\romannumeral #1@}

\input epsf
\usepackage{babel}\addto\captionsbreton{}
\addto\captionsbulgarian{}
\addto\captionsenglish{}

\usepackage{babel}\addto\captionsbreton{}
\addto\captionsbulgarian{}
\addto\captionsenglish{}

\usepackage{babel}\addto\captionsbreton{}
\addto\captionsbulgarian{}
\addto\captionsenglish{}

\@ifundefined{showcaptionsetup}{}{%
 \PassOptionsToPackage{caption=false}{subfig}}
\usepackage{subfig}
\makeatother

\usepackage{babel}
\begin{document}

\title{Analogue Radio Over Fiber for Next-Generation RAN: Challenges and Opportunities}
\author{
\IEEEauthorblockN{Yichuan Li, Qijie Xie, Mohammed El-Hajjar, \textit{Senior Member, IEEE}, Lajos Hanzo, \textit{Fellow, IEEE}}
\\

\vspace{-1cm}
\thanks{The financial support of of the China Postdoctoral Science Foundation
under Grant No. 2019TA0209, the EPSRC projects EP/N004558/1,
EP/L018659/1 as well as of the European Research Council's Advanced
Fellow Grant under the Beam-Me-Up project and of the Royal Society's
Wolfson Research Merit Award is gratefully acknowledged.}
\thanks{Yichuan Li is with Harbin Institute of Technology (Shenzhen), Shenzhen, China (e-mail: liyichuan@hit.edu.cn)}
\thanks{Qijie Xie is with Pengcheng Laboratory (xieqj@pcl.ac.cn)}
\thanks{M. El-Hajjar, and L. Hanzo are with the School of ECS,
University of Southampton, SO17 1BJ, United Kingdom ( meh@ecs.soton.ac.uk; lh@ecs.soton.ac.uk).}}
\maketitle
\begin{abstract}
The radio access network (RAN) connects the users to the core networks, where typically digitised radio over fiber (D-RoF) links are employed. The data rate of the RAN is limited by the hardware constraints of the D-RoF-based backhaul and fronthaul. In order to break this bottleneck, the potential of the analogue radio over fiber (A-RoF) based RAN techniques are critically appraised for employment in the next-generation systems, where increased-rate massive multiple-input-multiple-output (massive-MIMO) and millimeter wave (mmWave) techniques will be implemented. We demonstrate that huge bandwidth and power-consumption cost benefits may accrue upon using A-RoF for next-generation RANs. We provide an overview of the recent A-RoF research and a performance comparison of A-RoF and D-RoF, concluding with further insights on the future potential of A-RoF. 
\end{abstract}


\section{Introduction}
\label{Section:Introduction}
\begin{figure*}[h]
\centering
	\includegraphics[width=\textwidth]{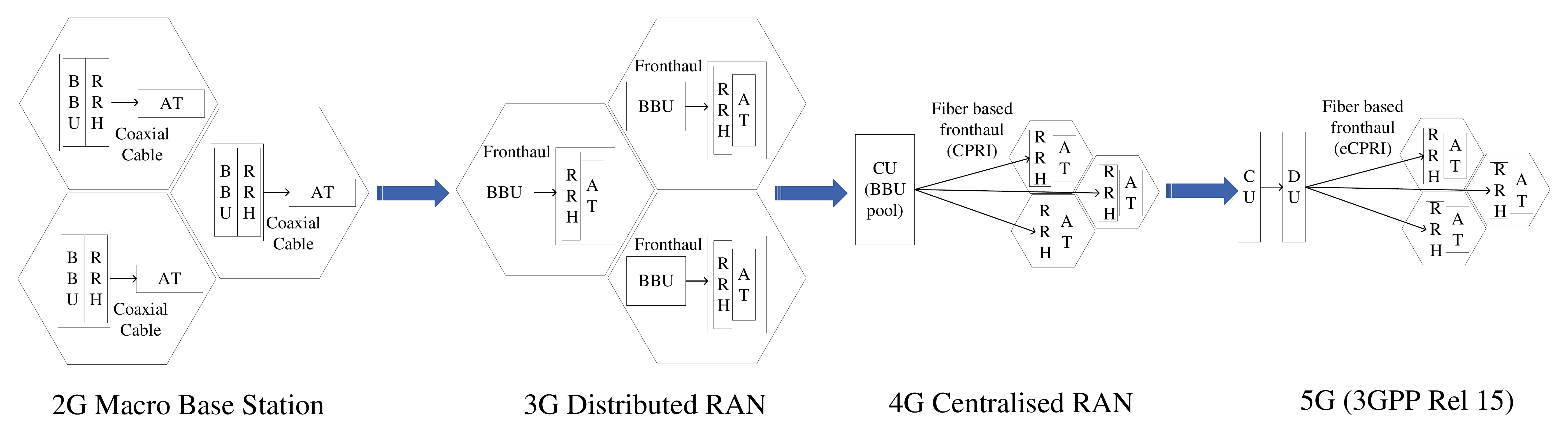}
		\caption{RAN Evolution (RRH: Remote Radio Head, BBU: Baseband Unit, AT: Antenna, CU: Central Unit, DU: Distributed Unit, eCPRI: enhanced Common Public Radio Interface.)}
	\label{Figure:RANEvolution}
\end{figure*} 

Radio access networks (RAN) have been evolving throughout the past five wireless generations. In the fourth Generation (4G) era, the RAN architecture typically relied on the concept of centralised RANs (C-RAN), where the central unit (CU) has several baseband units (BBUs) and can be connected to several remote radio heads (RRHs) by fiber, where each RRH supports an individual cell \cite{6897914}. The service-oriented fifth Generation (5G) evolved further by relocating some functions of the CU to the distributed unit (DU) for the sake of flexibly serving all use-cases, while some physical-layer functions were moved to the RRH connected by digitised radio over fiber (D-RoF) links \cite{osseiran20165g,5687951}.

As a mainstream 4G C-RAN fronthaul technique, D-RoF supports high-integrity data transmission by relying on sampled and quantized radio frequency (RF) signals over fiber, while dispensing with digital signal processing in the RRHs. However, following the emergence of high-rate massive multiple-input-multiple-output (massive-MIMO) and millimeter Wave (mmWave) communication techniques, the fully centralised D-RoF based 4G RAN architecture, where only radio functions are performed at the RRHs \cite{osseiran20165g}, requires hundreds of Gigabit per second (Gbps) data rates between the CU and the RRH.  

To meet these escalating capacity demands, the enhanced Common Public Radio Interface (eCPRI) design \cite{cpri2017ecpri} of Fig \ref{Figure:RANEvolution} can be employed for data transmission over the fiber using the existing D-RoF techniques by restricting the 5G fronthaul data rate below 30 Gb/s. However, this is achieved at the cost of increasing the complexity of RRHs \cite{cpri2017ecpri} by assigning them more complex tasks. Furthermore, the sampling and quantization process of D-RoF increases the bandwidth requirement of the fronthaul fiber-link in line with the quantizer's resolution. Hence the question arises whether analogue radio over fiber (A-RoF) might be able to circumvent this problem \cite{8295204,7017473} by dispensing with quantization all together, and if so-at what price. Indeed, there are some existing A-RoF implementations in the C-RAN systems, which we will elaborate on in Section \ref{Section:Status of A-RoF aided RAN}. These techniques are capable of supporting the 5G rate, but require further improvements for next-generation systems relying on increased-rate massive-MIMO and mmWave solutions due to the susceptibility of A-RoF to the fiber-links's nonlinearity and dispersion \cite{7017473}.

Against this backdrop, we present the design of fully centralised A-RoF aided C-RANs for supporting high-rate applications. Explicitly, we critically appraise the family of A-RoF aided RAN systems as a low-cost, low-power solution. Hence a suite of advanced fiber-link impairment mitigation techniques will be suggested in the context of high-bandwidth, high-order modulation based massive-MIMO solutions.

The rest of this paper is organised as follows. Section \ref{Section:A brief history of the Radio Access Network Evolution} provides a brief historical perspective on the radio access network's evolution. Then, we discuss both the bandwidth and the power consumption reduction of A-RoF compared to D-RoF in Sections \ref{Section:Bandwidth Efficiency of A-RoF over D-RoF} and \ref{Section:Power Consumption of A-RoF over D-RoF}, while the current status of A-RoF aided 5G applications is presented in Section \ref{Section:Status of A-RoF aided RAN}. Finally, we provide an outlook in Section \ref{Section: Future Research}, followed by our conclusions in Section \ref{Section: conclusion}.
\begin{figure*}[h]
\centering
	\includegraphics[width=.7\textwidth]{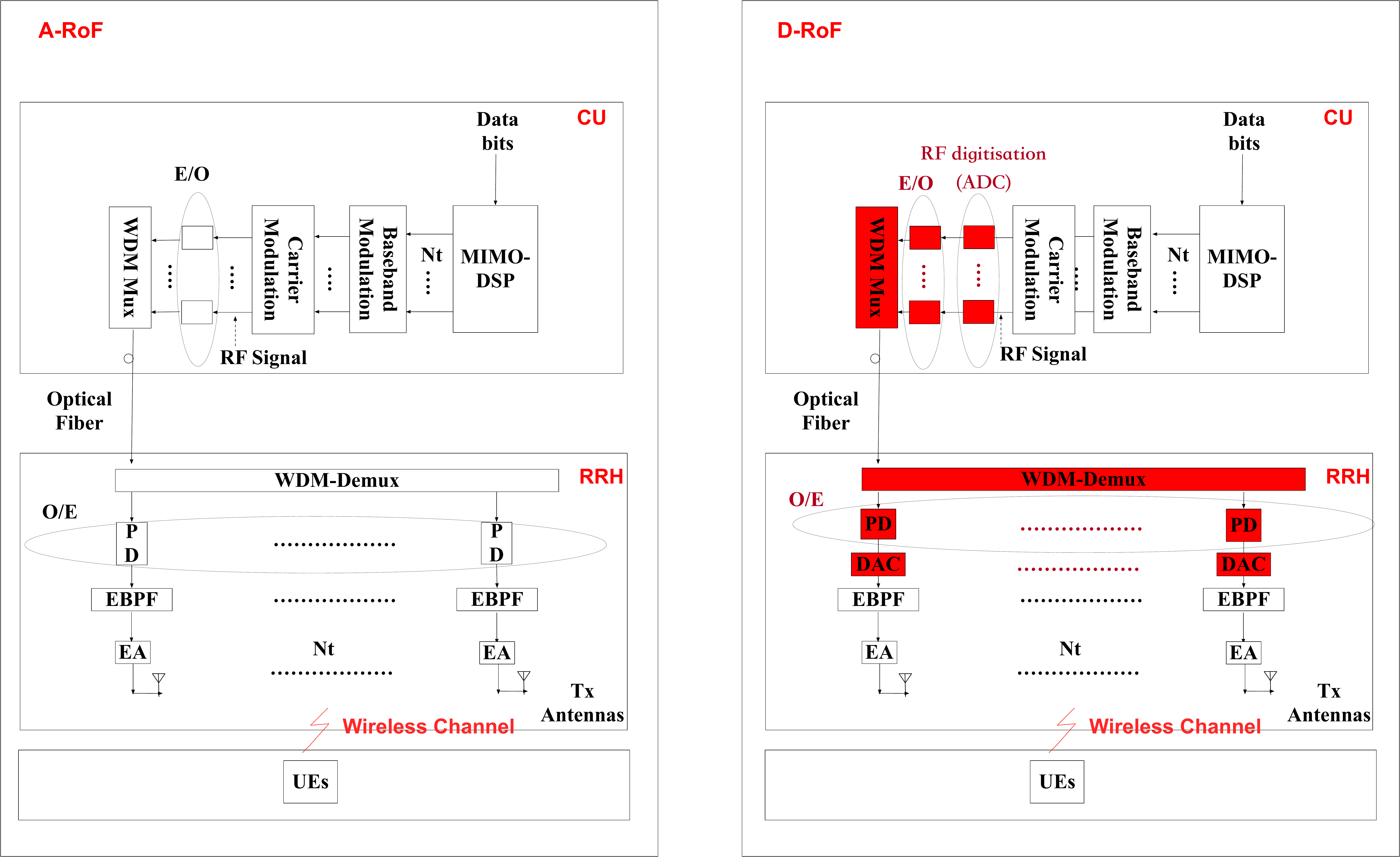}
		\caption{Left: A-RoF, Right: D-RoF. (DSP: Digital Signal Processing, EBPF: Electronic Bandpass Filter, EA: Electronic Amplifier, UE: User End.)}
	\label{Figure:RoF}
\end{figure*}
\section{A brief history of the Radio Access Network Evolution}
\label{Section:A brief history of the Radio Access Network Evolution}
As shown in Fig. \ref{Figure:RANEvolution}, the 2G cellular network relied on the macro base-station (BS) concept, where all the equipment except for the antennas was placed in a cell tower, with the copper-connected antennas being placed at the top of each tower \cite{6897914}. Then, to reduce the RF transmission attenuation, the radio unit or RRH was moved to be near the antenna in the distributed 3G RAN, where separating the radio- and baseband-functions led to the concept of fronthaul networks, as shown in Fig. \ref{Figure:RANEvolution} \cite{6897914}. Subsequently, in the 4G era, the C-RAN concept was born, where a single CU accommodating several baseband units is connected to multiple RRH by fiber. This architecture substantially reduces the operational expenditure in terms of the management and cooling cost, due to the small size of each cell tower \cite{6897914}. The C-RAN architecture utilised D-RoF techniques in the fronthaul network for the sake of eliminating most of the electronic processing in the RRHs, such as MIMO processing and up-conversion. Subsequently, the 5G RAN at the right of Fig. \ref{Figure:RANEvolution} adopted the functional split concept, where some electronic functions such as modulation are moved to the RRH to relieve the bandwidth requirement imposed on the fiber-link fronthaul, where the fronthaul standard employed is the eCPRI. The evolution of the RAN has been governed by its bandwidth requirement, especially in the context of massive-MIMO and mmWave communications. As mentioned in Section \ref{Section:Introduction}, A-RoF has a much lower bandwidth requirement and power consumption of the D-RoF based fronthaul, because the latter expands the bandwidth in line with the number of quantization bits. Hence in this paper we critically appraise the potential of A-RoF as a compelling RAN solution \cite{osseiran20165g,5687951} by comparing the A-RoF and D-RoF both from a bandwidth-efficiency and power-consumption perspective.
\section{A-RoF versus D-RoF: Bandwidth Efficiency Analysis}
\label{Section:Bandwidth Efficiency of A-RoF over D-RoF} 
In Fig. \ref{Figure:RoF}, we compare the A-RoF based C-RAN system to its D-RoF counterpart, when transmitting a $N_t \times N_r$ MIMO signal, with $N_t$ and $N_r$ representing the number of transmit and receive antennas. In the A-RoF, no analogue-to-digital conversion (ADC) and digital-to-analogue conversion (DAC) is required. In the A-RoF's CU, the data bits are modulated and subjected to MIMO processing, followed by up-conversion. The resultant signal is then electronic-to-optical (E/O) converted, as seen in Fig. \ref{Figure:RoF}. The wavelength-division-multiplexing multiplexer (WDM-mux) generates a WDM signal \cite{agrawal2012fiber}, where each wavelength carries its individual RF signal over the optical fiber to the RRHs of Fig. \ref{Figure:RoF}. In the A-RoF's RRH of Fig. \ref{Figure:RoF}, the WDM signal is fed into the WDM demultiplexer (WDM-demux). Then, each wavelength is optical-to-electronic (O/E) converted for recovering the RF signals using photo-detectors (PD). Finally, after bandpass filtering and electronic amplification, the RF signals carried by each wavelength of the WDM signal are mapped to each antenna for wireless MIMO transmission, where the MIMO signal processing was carried out in the CU. By contrast, in D-RoF, the RF signal is sampled and quantized prior to the E/O conversion, where the resultant digital signal is carried by the optical signal over the fiber. More explicitly, the bandwidth of the D-RoF signal is expanded by the sampling and quantization of the RF signal \cite{7017473}.
\subsection{Bandwidth Saving by A-RoF}
\label{SubSection:Bandwidth Compression}
Consider a MIMO system employing $N_t$ transmit antennas (TA) and $N_r$ receive antennas (RA) implemented in each 360$^{\circ}/M$ wide sector of each RRH, where the number of sectors is $M$. 

In the case of A-RoF, we assume that the baseband bandwidth is $W_{bb}$. Then, upon assuming that each TA has its own dedicated WDM carrier and each of the $M$ sectors also has the same bandwidth per carrier, we have a double-sided bandwidth of $2W_{bb}$ after RF carrier modulation, yielding a total A-RoF bandwidth of $W=2W_{bb}N_{t}M$. In other words, the aggregated bandwidth represents the total bandwidth of all the wavelength $N_t$ and of all the sectors $M$ involved in a WDM system. Furthermore, according to Nyquist's theorem \cite{7009970}, the corresponding A-RoF bit rate would be $2W_{bb}N_{t}M\log_2(S)$, where $S$ is the number of modulation constellation points \cite{tornatore2017fiber}.

In the case of D-RoF systems, when bandpass sampling is invoked \cite{tornatore2017fiber}, the sampling rate $f_s$ can be expressed as follows \cite{7009970}:
\begin{equation}
\begin{array}{lcl}
(2f_{max})/n_z\ \le f_{s}\le(2f_{min})/(n_z-1),\label{Equation5}
\end{array}
\end{equation}
where $1\le n_z\le |f_{max}/(f_{max}-f_{min}\ )| $ and $f_{max}=f_c+W_{bb}$ represents the maximum frequency edge of the RF signal, while $f_{min}=f_c-W_{bb}$ is its minimum frequency edge and $f_c$ is the carrier frequency.

To elaborate further on the D-RoF system, the minimum new bit rate after the resampling and quantization of the RF signal becomes $D=f_{s}RN_{t}M(2)C_wC$ \cite{7009970}, where $R$ is the quantizer's sampling resolution, the factor 2 is the multiplicative factor for the I/Q data, $C_{w}$ is the control signalling rate carrying the control and management information and C is the coding rate of the line coding employed. Then, if perfect rectangular pulse shaping is used, the D-RoF bandwidth is $D/2$.

As a result, we arrive at the bandwidth ratio $B$ and the bit rate ratio $C$ between the D-RoF and A-RoF as:
\begin{equation}
\begin{split}
\begin{array}{lcl}
B=\frac{D/2}{W}=\frac{f_{s}RN_{t}M(2)C_wC/2}{2W_{bb}N_{t}M}
=\frac{f_{s}RC_wC}{2W_{bb}} \\ \text{and}\\
C=\frac{D}{2W_{bb}N_{t}M\log_2(S)}=\frac{f_{s}RN_{t}M(2)C_wC}{2W_{bb}N_{t}M\log_2(S)}
=\frac{f_{s}RC_wC}{W_{bb}\log_2(S)}.
\end{array}
\end{split}
\label{Equation4}
\end{equation}

Next, we report on an experiment, in which a stream of 256 quadrature amplitude modulated (QAM) symbols are carried by a 28 GHz RF carrier supporting a $N_t=16$ MIMO system, where the number of 120$^\circ$ sectors per cell is $M=3$. Furthermore, we used a sampling resolution of $R=15$ for D-RoF digitisation, as well as a line-coding rate of $C_w=16/15$ and $C=10/8$, which follow the CPRI-based fronthaul specification \cite{cpri2017ecpri,7402275}. 

\begin{figure}[h]
\centering
	\includegraphics[width=.5\textwidth]{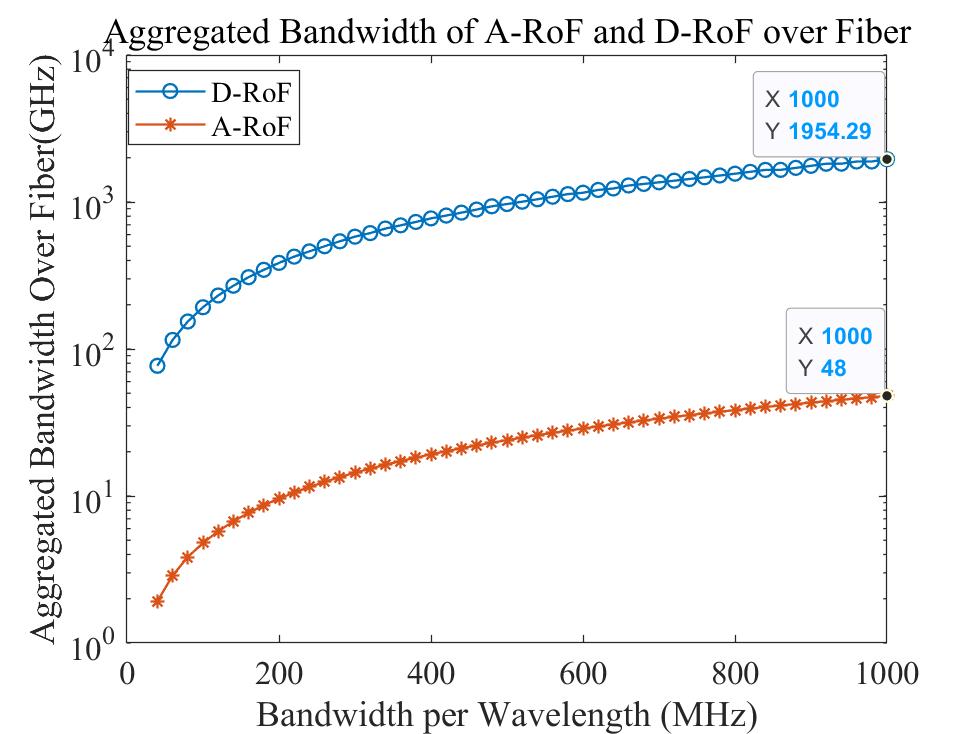}
		\caption{Aggregated Bandwidth of A-RoF and D-RoF over Fiber}
	\label{Figure:Bandwidth Ratio}
\end{figure}

In Fig. \ref{Figure:Bandwidth Ratio}, we show the data rate transmitted both by A-RoF and D-RoF, when the bandwidth per wavelength ranges from 50 MHz to 1 GHz. As shown in Fig. \ref{Figure:Bandwidth Ratio}, the achievable data rate of the A-RoF system may be a factor of 40 higher than that of the D-RoF system having the same optical bandwidth. Conversely, A-RoF may require a factor of 40 lower optical bandwidth for the same data rate, where the bandwidth per wavelength is 1 GHz.

To elaborate a little further, Fig. \ref{Figure:Bandwidth Ratio} was generated from Eq. (\ref{Equation4}) by gradually increasing the bandwidth of each wavelength from 20 MHz to 1 GHz. As shown in Fig. \ref{Figure:Bandwidth Ratio}, upon using a bandwidth per wavelength of 1 GHz, we obtain the minimum sampling rate $f_s$ as 2.0357 from Eq. (\ref{Equation5}). Thus, by setting $f_s$ according to Eq. (\ref{Equation4}), the aggregated D-RoF bandwidth increases to 1954 GHz, while only 48 GHz is required by A-RoF. Explicitly, their ratio is about 40, when both transmit at a 384 Gbps throughput, as dictated by $2W_{bb}N_{t}M\log_2(S)$. 

\subsection{Analysis of the Industrial D-RoF Market}
\label{Subsection:Impact on D-RoF Market}
Given the above bandwidth savings of the A-RoF fronthaul over its D-RoF counterpart, in this section, we review the practical implementation of the above-mentioned D-RoF system by considering state-of-the-art commercial products, in order to further demonstrate the potential benefits of A-RoF in next-generation RANs.
As seen in Fig. \ref{Figure:RoF}, D-RoF requires high-specification and costly E/O, WDM-mux, WDM-demux, and PD components capable of processing a high bandwidth. Hence, when considering the A-RoF bandwidth per wavelength of 1 GHz (8 Gbps per wavelength) using the configuration of Section \ref{SubSection:Bandwidth Compression}, its D-RoF bandwidth carried by each wavelength relying on digitisation would be 40.71 GHz (81.43 Gbps per wavelength). However, there is a paucity of both commercial E/O converters supporting a bandwidth of 40.71 GHz, and WDM Mux/Demux operating at 81.43 Gbps per wavelength. Furthermore, carrying an RF signal of an aggregated bandwidth of 1954 GHz over a single mode fiber would be exceptionally challenging due to the severe fiber nonlinearity and fiber dispersion imposed by a high-bandwidth and high-power system \cite{agrawal2012fiber}.

In addition to its lower bandwidth requirement, an A-RoF based fronthaul requires less power due to its simplified RRH configuration. Next, we focus on our attention the power consumption aspects.
\section{A-RoF versus D-RoF: Power Consumption Analysis}
\label{Section:Power Consumption of A-RoF over D-RoF} 
In this section, the power consumption of the A-RoF and the D-RoF system of Fig. \ref{Figure:RoF} is analysed using the model proposed by Jung \textit{et al.} \cite{7022990}, where the source of the total power consumption dissipation is detailed. According to \cite{7022990}, the power consumption of the CU is imposed by the climate control unit ($P_{CC}$), alternative circuit (AC)/ direct circuit (DC) power supply unit ($P_{ps\_CU}$), the CU's signal processing unit ($P_{sp\_CU}$) including the MIMO DSP, the baseband modulation, the carrier modulation (and ADC for the D-RoF) and the E/O convertor ($P_{eo}$). By contrast, the power dissipation of the RRH is due to the O/E convertor ($P_{oe}$), the AC/DC power supply unit ($P_{ps\_RRH}$) and the RRH signal processing unit ($P_{sp\_RRH}$), including the electronic band-pass filter (EBPF) (and the DAC for the D-RoF). 

\begin{figure}[h]
\centering
	\includegraphics[width=.5\textwidth]{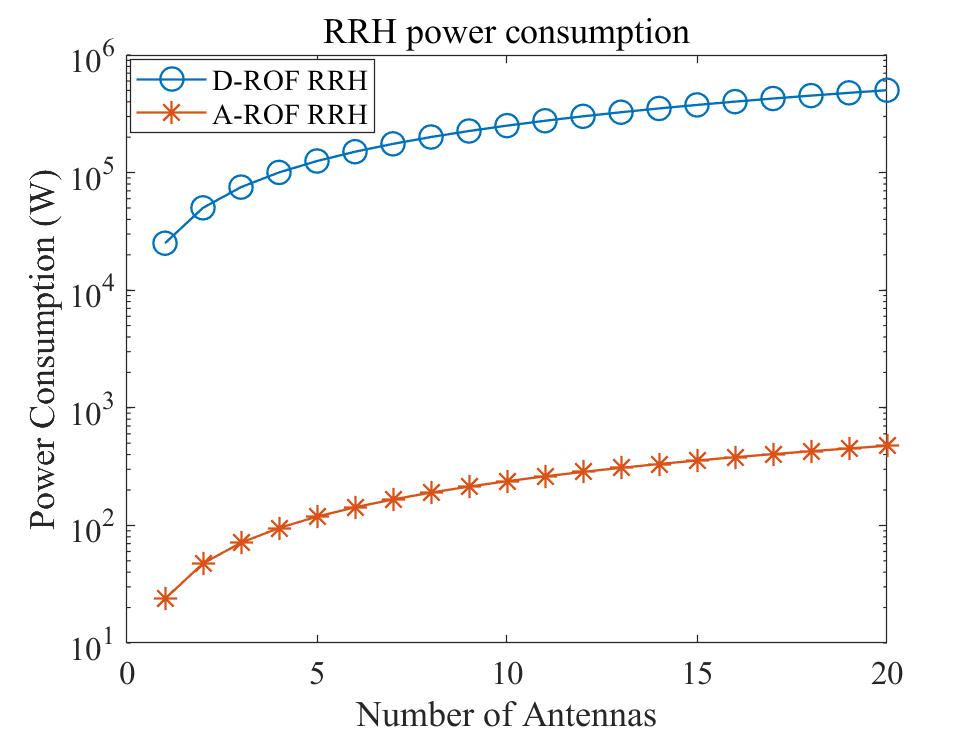}
		\caption{Power Consumption of A-RoF and D-RoF over Fiber}
	\label{Figure:AntennaRRHPowerConsumption}
\end{figure}

Note that in the RRH of Fig. \ref{Figure:RoF} and as analysed in \cite{7022990}, where the parameter values are presented. $P_{sp\_RRH}$ of the A-RoF and D-RoF would be different due to the DAC of the D-RoF, resulting in a different signal processing power consumption. Thus, based on an A-RoF bit rate of 20 Gbps, which would correspond to about 200 Gbps digital signalling over the fiber for D-RoF, Fig. \ref{Figure:AntennaRRHPowerConsumption} shows that A-RoF is significantly more energy efficient than its D-RoF counterpart. 

However, there are several challenges that the A-RoF faces, which are highlighted in the following section, followed by some promising research directions.
\begin{figure*}[h]
\centering
\includegraphics[width=\textwidth]{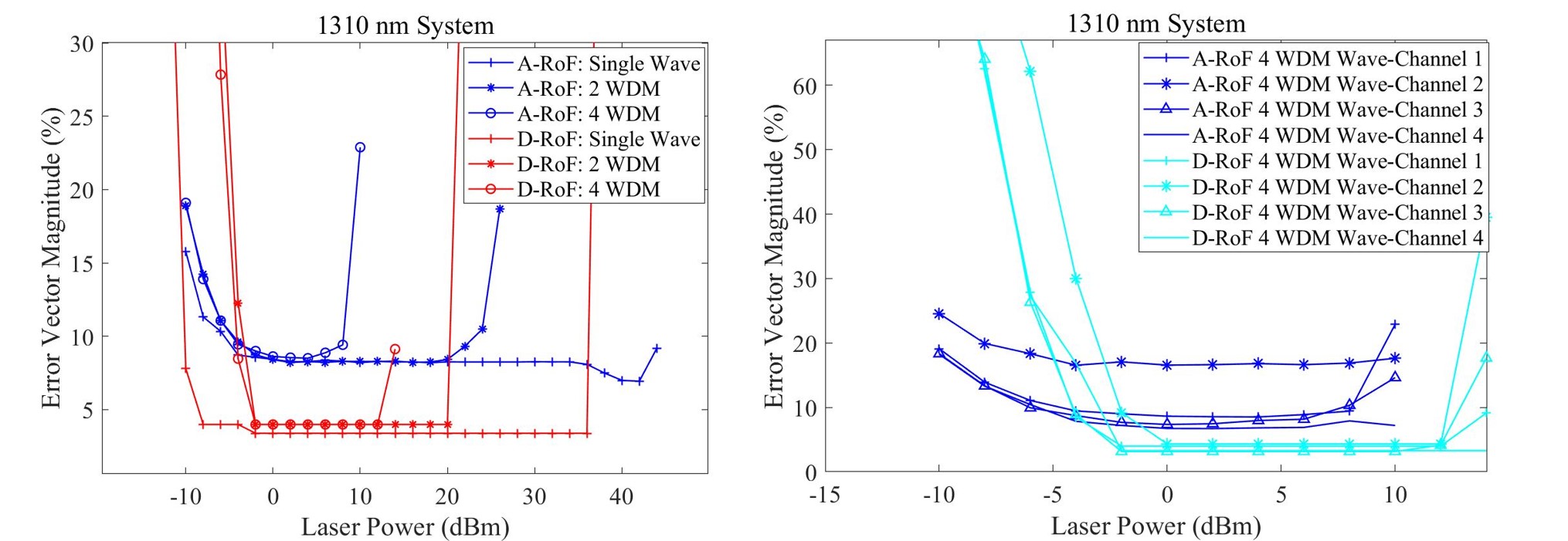}
\caption{EVM of A-RoF versus D-RoF}
\label{Figure:AntennaVSEVM}
\end{figure*}
\color{blue}
\begin{table*}[h]
		\begin{center}
		\scalebox{.8}{
		\begin{tabular}{lp{3cm}p{2cm}p{2cm}lll}
		\hline
		\hline Fronthaul Technique & Bandwidth Efficiency & Power & Performance & 5G EVM with 64-QAM & 5G EVM with 256-QAM & Standard\\
		    \hline A-RoF & up to 40-fold better than D-RoF &  Power-efficient with massive MIMO & Poor EVM performance and robustness & Satisfied & not Satisfied & None \\
		    \hline D-RoF & Bandwidth-thirsty & Power-thirsty with massive MIMO & High EVM performance and robustness & Satisfied & Satisfied & eCPRI\\	   
		    \hline
		    \hline
		\end{tabular}
		}
	\end{center}
	\caption{A comparison of A-RoF and D-RoF Fronthaul}
		\label{table:T1}
	\end{table*}
\color{black}
\section{Status of A-RoF aided RAN}
\label{Section:Status of A-RoF aided RAN}
As discussed in Section \ref{Section:Bandwidth Efficiency of A-RoF over D-RoF}, the A-RoF technique is capable of reducing the bandwidth of D-RoF by a factor of $\frac{f_{s}RC_wC}{2W_{bb}}$, which is as high as 40 upon using 256-QAM for 16$\times$16 MIMO transmission. In this section, a brief survey is given for reflecting the current status of A-RoF solutions, followed by the performance comparison of A-RoF over D-RoF using high-order modulation based high-bandwidth C-RAN systems.
\subsection{Current Status}
\label{subsection: Current Status}
Given that A-RoF is more susceptible to the fiber nonlinearity than D-RoF, to the fiber's chromatic dispersion and to optical noise as well as to the frequency chirp of directly modulated laser or E/O's nonlinearity \cite{7017473}, a comprehensive survey of their mitigation techniques was presented in \cite{8496890,7017473} by considering the A-RoF's fiber-link impairments. In this section, we provide an update on popular techniques of mitigating fiber nonlinearity, the E/O's nonlinearity and fiber dispersion.

\textbf{Fiber-nonlinearity:} Fiber-nonlinearity is a power-dependent impairment, which generates upper harmonics and also non-linear phase rotation. As a result, when WDM carries multiple wavelengths over a single fiber, the optical power of each wavelength affects the phase of the optical signals due to the power-dependent optical Kerr effect \cite{7009970} caused by the fiber-nonlinearity. For the traditional single-wave based RAN, the effect of fiber-nonlinearity is relatively modest due to the low optical power. However, WDM carrying multiple wavelengths at an increased power makes fiber-nonlinearity the key factor of influencing the system performance. There are several A-RoF nonlinearity mitigation techniques \cite{8633369,8552898}. For example, a mmWave A-RoF link relying on digital predistortion techniques was designed in \cite{8633369} for mitigating the fiber-nonlinearity, which achieved a 24 Gbps per wavelength throughput using 16-QAM without employing any digital components in the RRHs \cite{8633369}. Additionally, a 16-DWDM based A-RoF system having a bandwidth of 20 MHz per wavelength was proposed for a massive-MIMO system by mitigating the non-linear stimulated Brillouin scattering \cite{8552898}. 

\textbf{E/O's Nonlinearity: } The E/O's nonlinearity is caused by the high amplitude of its drive signal, which degrades the system performance owing to contamination by the resultant upper harmonics \cite{7017473}. For example, the externally modulated laser inflicts contamination by the upper harmonics onto its output, when its RF drive signal exhibits high power. Hence, directly modulated laser can be used for reducing the effect of nonlinearity. A 10-wavelength WDM-based A-RoF system has successfully transmitted both 16- and 64-QAM symbols over a 20-km fiber at a bandwidth of 2.5 GHz per wavelength \cite{8737735}. As a further development, the optimal RF power and optical power was found in \cite{8295204} by analysing the externally modulated laser's nonlinear curve, which facilitated reliable transmission by using a bandwidth of 2 GHz per wavelength over 20-km length of fiber \cite{8295204}. Thus, the A-RoF system exploiting either the directly modulated laser or the externally modulated laser is investigated for reducing the E/O's nonlinearity.

However, dispersion tends to become more severe in high-bandwidth communication, hence requiring sophisticated counter-measures \cite{8496890,7017473}.

In a nutshell, it may be concluded that similar to D-RoF, WDM-based A-RoF aided MIMOs also have their pros and cons, but they are capable of meeting industry's requirements as discussed below with reference to Table \ref{table:T1}. 
\begin{enumerate}
\item \textbf{Performance:} In exchange for its wide bandwidth requirement, D-RoF exhibits superior error vector magnitude (EVM) performance and a high degree of flexibility without customising the system parameters for various environments \cite{7017473,tornatore2017fiber}.
\item \textbf{Bandwidth and Bit Rate:} As mentioned in Section \ref{Section:Introduction} and Section \ref{Section:Bandwidth Efficiency of A-RoF over D-RoF}, D-RoF is capable of supporting the 20 Gbps 5G system target, while using off-the-shelf optical components. This corresponds to an aggregated digitised data rate of almost 200 Gbps and $<10$ Gbps per wavelength in a 20$\times$20 MIMO system, which can be readily realised by using commercial optical devices. Its comparison to A-RoF is shown in Table \ref{table:T1}.
\item \textbf{Standardisation:} The fronthaul market is based on the D-RoF CPRI/eCPRI solution, but there is no standard for A-ROF system.
\end{enumerate}

Hence, again, in the 5G era D-RoF has better EVM performance than its A-RoF counterpart, which justifies the industry's preference for D-RoF. Table \ref{table:T1} compares them in terms of their bandwidth efficiency, power efficiency, EVM performance, 5G feasibility and standardisation status. Explicitly, the benefits of A-RoF are constituted by its attractive bandwidth and power efficiency, but its EVM performance has to be improved when using 256-QAM as discussed in the next section.

Quantitatively, below we consider a 256-QAM system and a baseband bandwidth of 1.87 GHz for the performance comparison of A-RoF and D-RoF in the next section and highlight some of the challenges of the A-RoF fronthaul.
\subsection{An overview of A-RoF and D-RoF Performance in WDM-based C-RAN Scenario}
\label{subsection: An overview of A-RoF and D-RoF Performance in WDM-based C-RAN Scenario}
In this section, we focus our attention on the performance analysis of A-RoF versus D-RoF in a high-bandwidth and high-rate WDM-based C-RAN scenario using the MATLAB-Optisystem. The performance is evaluated using the EVM, a metric that quantifies the distortion of the received signal constellation for various number of MIMO antennas. We follow the typical C-RAN architecture of Fig. \ref{Figure:RoF}, where the RF signal is generated off-line using MATLAB. Naturally, A-RoF conveys analogue signals, while D-RoF transmits a digitised signal after the ADC. Then, our Optisystem simulates the E/O, WDM Mux/Demux, fiber transmission and PDs. The recovered RF signals are filtered, amplified and analysed by MATLAB. 

We consider a network transmitting at a rate of 20 Gbps per wavelength, which is carried by an RF signal of 3 GHz bandwidth, where 256-QAM and a standard single mode fiber (SSMF) of 15 km are used. The wavelength spacing of the WDM signal is 200 GHz, while the central wavelength is 1310 nm. To analyse the impact imposed by the number of antennas on the fiber nonlinearity, Figure \ref{Figure:AntennaVSEVM} shows the EVM performance of both A-RoF and D-RoF upon varying the number of antennas. 

Explicitly, observe in Fig. \ref{Figure:AntennaVSEVM} that the A-RoF system fails to meet the 3GPP EVM target of 3.5\% \cite{6897914}, while the D-RoF system exhibits an EVM of around 3.5\%, which just meets the 3GPP's requirement for transmitting 256-QAM symbols. By contrast, the A-RoF shows EVM$\approx$8\%, due to the analogue signal's susceptibility to fiber nonlinearity. Furthermore, upon increasing the WDM's dimension, the dynamic range of the A-RoF system (i.e. the range of the laser power under which the EVM remains within a certain limits) becomes poorer than that of D-RoF. Explicitly, it is around 8 dBm for the 4-WDM A-RoF system compared to 14 dBm for the 4-WDM D-RoF system. 

Furthermore, using this 4-WDM system as our example, the right graph of Fig. \ref{Figure:AntennaVSEVM} compares the EVM of the signals transmitted over the channels operating at different center wavelengths, where D-RoF substantially outperforms A-RoF.
\section{Future Research of A-RoF}
\label{Section: Future Research}
\begin{enumerate}
\item \textbf{Mitigating fiber-dispersion and fiber-nonlinearity of A-RoF in high-bandwidth scenarios:} 

~The employment of optical carrier and side-band suppression techniques as well as accurately biasing the optical modulators are expected to be helpful in terms of mitigating the fiber-dispersion \cite{7017473}. On the other hand, the A-RoF fronthaul operating at 1310 nm is capable of reducing the fiber's chromatic dispersion, hence potentially dispensing with the employment of high-complexity dispersion compensation techniques in the RRHs. However, the deleterious nonlinear effect of four-wave mixing may seriously erode the system's performance. Thus, a carefully balanced trade-off has to be maintained between the fiber-dispersion and fiber-nonlinearity.

\item \textbf{Mitigating the non-linear effects in high-order modulation and massive-MIMO scenarios:} 

~When considering a 256-QAM system, maximising the signal-to-noise ratio (SNR) is crucial. Increasing the optical power and the RF signal power would improve the SNR, but the increased optical power increases the impact of fiber-nonlinearity, which then results in increased E/O conversion nonlinearity. 

~Therefore, both the E/O converter's and the fiber's nonlinearity are affected both by the RF signal power and by the optical power. Hence, the joint optimization of both powers is essential for introducing both WDM and 256-QAM into A-RoF systems, which would have to be followed by verification using testbed implementations. 

~Additionally, the optical power of each wavelength of the WDM based massive-MIMO system may also affect the neighbouring wavelengths as a result of the fiber's nonlinearity, where each WDM channel experiences different EVM performance, as shown in the right graph of Fig. \ref{Figure:AntennaVSEVM}. For example, as shown in the left graph of Fig. \ref{Figure:AntennaVSEVM}, the dynamic range of the A-RoF system is degraded upon increasing the WDM dimension, indicating that the power of the neighbouring wavelength adversely affects the system performance. Thus, finding the optimal power levels of all the WDM wavelengths would benefit the A-RoF's robustness.
\item \textbf{Silicon Photonics for facilitating the A-RoF deployment:}

~The key driving force behind silicon photonics is the potential of adopting the existing Complementary Metal Oxide Semiconductor (CMOS) fabrication technology for photonic integration in order to create low-cost, yet high-performance systems. On one hand, the performance of A-RoF is directly dependent on the fiber-quality. Hence, the employment of silicon photonics may reduce the nonlinearity or noise of the photonic devices employed. On the other hand, the objective of both A-RoF and of silicon photonics is that of reducing the system's complexity. Since A-RoF requires lower bandwidth than D-RoF, having an A-RoF based silicon photonics architecture meeting the current capacity may be deemed realistic.
\end{enumerate}

Although there are critical challenges ahead, promoting the standardisation of A-RoF based RANs may inspire the industry to fabricate the required analogue components. Specific use-cases, such as small-cell cellular networks may be considered for the initial employment of A-RoF, since they require high-rate data transmission and small RRHs, where A-RoF has the edge in terms of its bandwidth- and energy-efficiency over D-RoF.
\section{Conclusions}
\label{Section: conclusion}
The operational D-RoF based RAN fronthaul requires an excessive bandwidth over the fiber-link. Hence, we have critically appraised the potential of A-RoF solutions in circumventing this impediment in the context of massive MIMO aided high-order QAM schemes. We have compared their bandwidth- and power-efficiency, demonstrating that A-RoF may be capable of reducing the bandwidth required by a factor of 40, and hence it is capable of relaxing the hardware specifications, while significantly reducing the power-consumption. However, A-RoF requires substantial further research in order to guarantee the bit rate requirements of next-generation applications. 


 \bibliographystyle{ieeetr}
\bibliography{ECS}
 
\begin{IEEEbiography}
{Yichuan Li(liyichuan@hit.edu.cn)} is an Assistant Professor with Harbin Institute of Technology (Shenzhen). He received his B.Sc. degree in Optics Information Science and Technology from China University of Petroleum (East China), Qingdao, China, in 2012, and earned his M.Sc. degree and Ph.D. degree in wireless communications from the University of Southampton, Southampton, UK., in 2014 and 2019. He was a research assistant in the Lightwave Communication Lab of the Chinese University of Hong Kong (CUHK) from July to October in 2017. His research is focused on the the radio over fiber for backhaul, fronthaul and indoor communication network. His research interests are millimeter wave over fiber, optical fiber aided analogue beamforming techniques, Multi-functional MIMO, mode division multiplexing in multimode fiber and fiber-based C-RAN system.
\end{IEEEbiography}
 \begin{IEEEbiography}
{Qijie Xie (xieqj@pcl.ac.cn)} received his Ph.D. degree in Electronic Engineering from The Chinese University of Hong Kong (CUHK) in 2018. Until June 2020, he had been a postdoctoral research associate in Prof. Chester Shu's group, CUHK. Now, he is working as an assistant researcher in Pengcheng Laboratory, Shenzhen, China.
\end{IEEEbiography}
\begin{IEEEbiography}{Mohammed El-Hajjar(meh@ecs.soton.ac.uk)} is an Associate Professor in the Department of Electronics and Computer Science in the University of Southampton. He received his PhD in Wireless Communications from the University of Southampton, UK in 2008. Following the PhD, he joined Imagination Technologies as a design engineer, where he worked on designing and developing Imagination's multi-standard communications platform, which resulted in several patents. He is the recipient of several academic awards and has published a Wiley-IEEE book and in excess of 80 journal and conference papers. Mohammed's research interests include the development of intelligent communications systems, energy-efficient transceiver design, MIMO, millimeter wave communications and Radio over fiber network design.
 
\end{IEEEbiography}
\begin{IEEEbiography}
{\bf Lajos Hanzo(lh@ecs.soton.ac.uk)} (http://www-mobile.ecs.soton.ac.uk)  (FIEEE'04, Fellow of the Royal
Academy of Engineering F(REng), of the IET and of EURASIP), received
his Master degree and Doctorate in 1976 and 1983, respectively from the
Technical University (TU) of Budapest. He was also awarded the Doctor
of Sciences (DSc) degree by the University of Southampton (2004) and
Honorary Doctorates by the TU of Budapest (2009) and by the University
of Edinburgh (2015). He is a Foreign Member of the Hungarian Academy
of Sciences and a former Editor-in-Chief of the IEEE Press. He has served
several terms as Governor of both IEEE ComSoc and of VTS. He has
published 1900+ contributions at IEEE Xplore, 19 Wiley-IEEE Press books
and has helped the fast-track career of 123 PhD students. Over 40 of them
are Professors at various stages of their careers in academia and many of
them are leading scientists in the wireless industry.
\end{IEEEbiography}
\end{document}